\documentclass[prd,floatfix,twocolumn,nofootinbib,
preprintnumbers]{revtex4}
\usepackage[latin1]{inputenc}
\usepackage{amsmath}
\usepackage{amsfonts}
\usepackage{hyperref}
\usepackage{amssymb}
\usepackage{graphicx}
\usepackage{dcolumn}
\usepackage{caption}
\usepackage{subfigure}
\usepackage{float}
\usepackage{bm}
\usepackage{latexsym}
\usepackage{epsfig}
\usepackage{graphicx,multirow}
\usepackage{color}

\newcommand{\cg}[6]{\mathcal{C}^{#1 #2}_{#3 #4 #5 #6}}

\newcommand{\ncap}{\hat{n}}       
\newcommand{\al}{\text{\large{$\alpha$}}}
\newcommand{\be}{\begin{equation}}
\newcommand{\ee}{\end{equation}}
\newcommand{\ba}{\begin{eqnarray}}
\newcommand{\ea}{\end{eqnarray}}
\newcommand{\nn}{\nonumber \\}

\begin{document}

\title{WMAP anomaly : Weak lensing in disguise}
\author{Aditya Rotti\footnote{aditya@iucaa.ernet.in}, Moumita
Aich\footnote{moumita@iucaa.ernet.in} and Tarun
Souradeep\footnote{tarun@iucaa.ernet.in}}
\affiliation{IUCAA, Post Bag 4, Ganeshkhind, Pune-411007, India}
\date{\today}
\begin{abstract}
Statistical isotropy (SI) has been one of the simplifying assumptions in
cosmological model building. Experiments like WMAP and PLANCK are attempting to
test this assumption by searching for specific signals in the Cosmic Microwave
Background (CMB) two point correlation function. Modifications to this
correlation function due to gravitational lensing by the large scale structure
(LSS) surrounding us have been ignored in this context. Gravitational lensing
will induce signals which mimic isotropy violation even in an isotropic
universe. The signal detected in the Bipolar Spherical Harmonic (BipoSH)
coefficients $A^{20}_{ll}$ by the WMAP team may be explained by accounting for
the lensing modifications to these coefficients. Further the difference in the
amplitude of the signal detected in the V-band and W-band maps can be explained
by accounting for the differences in the designed angular sensitivity of the
instrumental beams. The arguments presented in this article have crucial
implications for SI violation studies. Constraining SI violation will only be
possible by complementing CMB data sets with all sky measurements of the large
scale dark matter distribution. Till that time, the signal detected in the BipoSH coefficients from WMAP-7
could also be yet another suggested evidence of strong deviations from the standard $\Lambda$CDM 
cosmology based on homogeneous and isotropic FRW models.
\end{abstract}
\maketitle
The Cosmic Microwave Background (CMB) anisotropy measurements are one of the cleanest probes of cosmology.
The CMB full sky temperature anisotropy measurements have been used to test the assumption
of the isotropy of the universe. Ever since the release of first year data of the
Wilkinson Microwave Anisotropy Probe (WMAP), statistical isotropy of the CMB
anisotropy has attracted considerable attention. The study of full sky maps from the
WMAP 5 year data \cite{komatsu,spergel,nolta} and the very recent WMAP 7 year
data \cite{CB-RH-GH}, has led to some intriguing anomalies that may be
interpreted to indicate deviations from statistical isotropy.

The CMB temperature anisotropies are assumed to be Gaussian, which is in good
agreement with current CMB observations.
Hence the two point correlation function contains complete
information about the underlying CMB temperature field. The two point
correlation function can be expressed in terms of the spherical harmonic
coefficients of CMB temperature maps,
\ba
	C(\mathbf{\ncap_1},\mathbf{\ncap_2})&=& \langle\Delta
T(\mathbf{\ncap_1})\Delta
T(\mathbf{\ncap_2})\rangle \\ 
	&=& \sum_{lml'm'} \langle a_{lm}a^*_{l'm'}\rangle Y_{lm}(\mathbf{\ncap_1})
Y_{l'm'}^{*}(\mathbf{\ncap_2})	\,.  \nonumber
\ea
In the case of statistical isotropy, the correlation function depends only on the
angular separation between the two directions and not on the directions
$\mathbf{\ncap_1}$ and $\mathbf{\ncap_2}$ explicitly. This property makes it
possible to expand the correlation function in the Legendre polynomial ($P_l$)
basis,
\ba
	C(\mathbf{\ncap_1},\mathbf{\ncap_2})&=&C(\mathbf{\ncap_1} \cdot
\mathbf{\ncap_2}) \\
	&=& \sum_l \frac{2l+1}{4\pi} C_{l} P_{l}(\mathbf{\ncap_1} \cdot
\mathbf{\ncap_2}) \,, \nn
	C_l&=& \langle a_{lm}a^*_{l'm'}\rangle \delta_{l l'}\delta_{mm'} \,.
\label{unlensed-cl}
\ea
The angular power spectrum, $C_l$, appearing  in Eq. \ref{unlensed-cl} encodes
all information in the covariance matrix 
for the statistically isotropic case. This however is only true in the case of
an unlensed CMB sky as will be evident from the primary message of the article.

In the absence of statistical isotropy, the correlation function explicitly
depends on the two directions $\mathbf{\ncap_1}$ and $\mathbf{\ncap_2} $. In
this case, the covariance matrix has been shown to have non-vanishing off-diagonal elements.
This feature of the covariance matrix is captured well in the Bipolar Spherical
Harmonic (BipoSH) representation which was introduced by Hajian \& Souradeep (HS)
\cite{AH-TS,AH-TS1}. The BipoSH form a complete orthonormal basis for functions defined on $S^2 \times
S^2$. The CMB two point correlation function can be expanded in the BipoSH basis
in the following manner,
\ba
	 C&
&\!\!\!\!\!\!\!\!\!\!\!\!(\mathbf{\ncap_1},\mathbf{\ncap_2})=\sum_{LMl_1l_2}
A^{LM}_{l_1 l_2} \left\{ Y_{l_1}
(\mathbf{\ncap_1})\otimes Y_{l_2}(\mathbf{\ncap_2})\right\}_{LM}\\
	&=&\sum_{L M l_1 l_2} A^{LM}_{l_1 l_2} \sum_{m_1 m_2} \cg L M {l_1}
{m_1} {l_2} {m_2} Y_{l_1m_1} (\mathbf{\ncap_1})Y_{l_2 m_2}(\mathbf{\ncap_2})\,,
\nonumber
\ea
where $\cg L M {l_1} {m_1} {l_2} {m_2}$ are the Clebsch-Gordon coefficients, the
indices of which satisfy the following relations, $|l_1-l_2|\leq L \leq l_1+l_2
$ and $m_1+m_2 = M$.

These BipoSH coefficients can be expressed in terms of the covariance matrix
derived from observed CMB maps,
\ba
 A^{LM}_{l_1 l_2} = \sum_{m_1 m_2} \langle a_{l_1 m_1}a_{l_2 m_2} \rangle \cg L
M {l_1} {m_1} {l_2} {m_2} \,.
\ea 
In the case of statistical isotropy the only non-vanishing BipoSH coefficients
are $A^{00}_{ll}$ and they can be expressed in terms of the CMB angular
power spectrum $C_l$,
\be \label{iBipoSH}
	A^{00}_{ll}=(-1)^l \Pi_l C_l, 	
\ee
where $\Pi_l=\sqrt{2l+1}$. We reiterate that the above discussion is
true only for the case of unlensed statistically isotropic CMB temperature
fields.

Weak lensing of the CMB photons due to scalar perturbations leave measurable
imprints on the CMB two point correlation function. The lensing modification to
the CMB angular power spectra have been well studied \cite{US1,WH}. 
It is well known that lensing introduces coupling between different multipole
moments of the temperature field, 
which otherwise are not expected to be present in an unlensed statistically
isotropic
 CMB sky (See Eq. \ref{unlensed-cl}). This coupling between the various
multipole moments 
arises due to the coupling between the projected lensing potential $\Psi(\ncap)$
 \cite{AL-AC} and the unlensed CMB temperature field. This feature again can be
most generally captured
 in the BipoSH representation.

In the discussion that follows we have assumed that the unlensed CMB temperature
field is statistically isotropic and that the projected lensing potential and
the CMB
temperature field are uncorrelated. Lensing remaps the temperature field. To
leading order terms in the deflection field $\mathbf{\Delta}$, the lensed
temperature field $\tilde{T}(\mathbf{\ncap})$ can be expressed in terms of the
unlensed temperature field $T$,
\ba
\tilde{T}(\mathbf{\ncap})&=&T(\mathbf{\ncap}+\mathbf{\Delta}) \, \nn
&\approx &T+\Delta^a \nabla_a T+ \frac{1}{2} \Delta^a \Delta^b \nabla_a \nabla_b
T \,,
\ea
where the lensing deflection field on the sky can be expressed in terms of the gradient of the
projected lensing potential field on the sphere as, 
\be 
\Delta_a=\nabla_a \Psi(\mathbf{\ncap}).
\ee
The two point correlation of the lensed temperature field can be expressed as
follows,
\ba \label{2ptcorr}
	\langle \tilde{T}(\mathbf{\ncap_1})\tilde{T}(\mathbf{\ncap_2})\rangle
&=&\langle
T(\mathbf{\ncap_1}+\mathbf{\Delta_1})T(\mathbf{\ncap_2}+\mathbf{\Delta_2}
)\rangle \nn
	&=& \langle T(\mathbf{\ncap_1})T(\mathbf{\ncap_2})\rangle \\
	&+& \langle\nabla^a \psi(\mathbf{\ncap_1}) \nabla_a
T(\mathbf{\ncap_1})T(\mathbf{\ncap_2})\rangle
\nn
	&+&\langle\nabla^a \psi(\mathbf{\ncap_2}) \nabla_a
T(\mathbf{\ncap_2})T(\mathbf{\ncap_1})\rangle +
O(\psi^2)\,.\nonumber
\ea
The corrections to the CMB angular power spectrum due to lensing arise only due
to
terms which are of $O(\psi^2)$, whereas the corrections to the BipoSH
coefficients can be shown to be only due to terms linear in the lensing
deflection field \cite{LB-MK-TS,AR-MA-TS}.

Without making assumptions about the isotropy of the lensed temperature field,
the two point correlation can be most generally expressed in terms of the BipoSH
coefficients and the spherical
harmonic coefficients of the lensing deflection field $\psi_{lm}$,
\ba
	\tilde{A}^{LM}_{l_1 l_2}=A^{LM}_{l_1 l_2}&+&{}_1\al^{LM}_{l_1
l_2}(\psi_{lm},A^{L'M'}_{l_1' l_2'})\nn
	&+&{}_2\al^{LM}_{l_1 l_2}(\psi_{lm},A^{L'M'}_{l'_1 l'_2})\,.
\ea
Since we have already assumed the unlensed temperature field to be
isotropic, the above equation can be further simplified and expressed (Using Eq.
\ref{iBipoSH} )  in terms of the unlensed CMB angular power spectrum,
\ba
	\tilde{A}^{LM}_{l_1 l_2}=C_{l_1} \delta_{l_1 l_2}\delta_{L0}\delta_{M
0}&+&{}_1\al^{LM}_{l_1 l_2}(\psi_{lm},C_{l_1'},C_{l_2'})\nn
	&+&{}_2\al^{LM}_{l_1 l_2}(\psi_{lm},C_{l_1'},C_{l_2'})\,.
\ea
Evaluating Eq. \ref{2ptcorr} in the harmonic space allows us to obtain the
following
expression for the coefficient $\al^{LM}_{l_1 l_2}$,
\ba \label{lensbips}
\al^{LM}_{l_1 l_2}&=&{}_1\al^{LM}_{l_1 l_2}+{}_2\al^{LM}_{l_1 l_2} \nn
&=& \psi_{LM} \frac{C_{l_1}F(l_2,L,l_1)+C_{l_2}F(l_1,L,l_2)}{\sqrt{4\pi}} \nn
& \times & \frac{\Pi_{l_1} \Pi_{l_2}}{ \Pi_{L}}~\cg L 0 {l_1} 0 {l_2} 0 \,,
\ea
where,
\ba
F(l_1,L,l_2)&=&\frac{\left[ l_2(l_2+1)+L(L+1)-l_1(l_1+1)\right]}{2} \,.
\nonumber
\ea
Note that lensing by scalar density perturbations necessarily generates only 
the even parity (i.e. $l_1+l_2+L$ is even)  BipoSH coefficients which is
apparent from Eq. \ref{lensbips} due to the presence
of $\cg L 0 {l_1} 0 {l_2} 0$ which vanishes for odd parity ($l_1+l_2+L$ is odd).

The above discussion of BipoSH coefficients is in terms of the HS estimator
which differs from the estimator used by the WMAP team. The WMAP estimator is
valid only for even parity Bipolar coefficients unlike the HS estimator. For even parity coefficients, the two
estimators are related by the following expression,
\be \label{WMAP-HS}
(A^{LM}_{l_1 l_2})_{\mathrm{WMAP}}=\frac{ \Pi_{L}}{\Pi_{l_1} \Pi_{l_2}}
\frac{1}{\cg L 0 {l_1} 0 {l_2} 0} (A^{LM}_{l_1 l_2})_{HS} \,.
\ee
In discussions that follow we use the WMAP estimator for the BipoSH
coefficients.
\begin{figure} [ht]
\includegraphics[height=8cm,width=5cm, angle=-90]{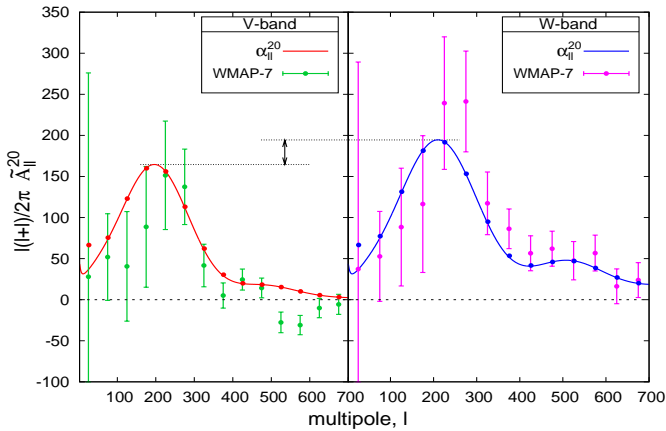}
\captionsetup{singlelinecheck=off,justification=raggedright}
\vskip 0.3cm
\captionof{figure}{The amplitude of the signal detected in the BipoSH
coefficient $A^{20}_{l l}$ in the V-band map is less than the amplitude
of the signal in the W-band map exactly corresponding to the expected difference due to the two instrumental beam widths.
It is remarkable and intriguing that it is possible to explain the V-band and W-band detections with a common, consistent value for the quadrupolar component of the projected lensing
potential $\psi_{20}$. See Table \ref{table1} for the beam specifications of
WMAP and the best fit value of the parameter $\psi_{20}$.}
\label{fig1}
\end{figure}

Motivated by the WMAP detections of isotropy violation \cite{CB-RH-GH} in the
V-band and W-band maps in ecliptic coordinates, we study the corresponding
BipoSH coefficients $A^{20}_{ll}$ and $A^{20}_{l,l+2}$ that arise due to
lensing. These coefficients take up an extremely simple form proportional to the angular power spectrum given by,
\begin{subequations}\label{cute2}
\ba 
\tilde{A}^{20}_{ll}~~&=&\al^{20}_{ll}=\frac{3 \psi_{20}
}{\sqrt{\pi}}C_{l}W^2_l,\label{cute} \\
\tilde{A}^{20}_{l,l+2}&=&\al^{20}_{l,l+2} \, \nn
&=&\frac{\psi_{20} }{\sqrt{\pi}}
\left[(l+3)C_{l+2}W^2_{l+2}-lC_lW^2_l\right]\label{cute1},
\ea
\end{subequations}
where $W^2_l$ corrects the CMB angular power spectrum $C_l$ for convolution of the instrumental response
beam function. 

We repeat the WMAP analysis to measure the BipoSH coefficients from the measured
V-band and W-band maps. In our analysis we account for the effects of the
anisotropic noise, instrument beam, foregrounds  and masking, each of which can
mimic signals of isotropy violation. We estimate the bias introduced due to all
the above mentioned effects (except for the non-circularity of the instrument beam) through
simulations. This bias is then subtracted from the signal inferred from the
analysis of 
the WMAP V-band and W-band maps to isolate the systematic effects. We do not
perform the inverse noise weighting that results in a much larger estimate for
the errors here and lowers the significance of the detections in our analysis. The broad features of
the detections are however retained and suffice for the arguments presented in
this article. More elaborate analysis incorporating 
inverse noise weighting on harmonic coefficients $a_{lm}$ of the temperature
field and the non-circularity of the beam is underway and will be reported in a more detailed publication.
\begin{figure} [ht]
\includegraphics[height=8cm,width=5cm, angle=-90]{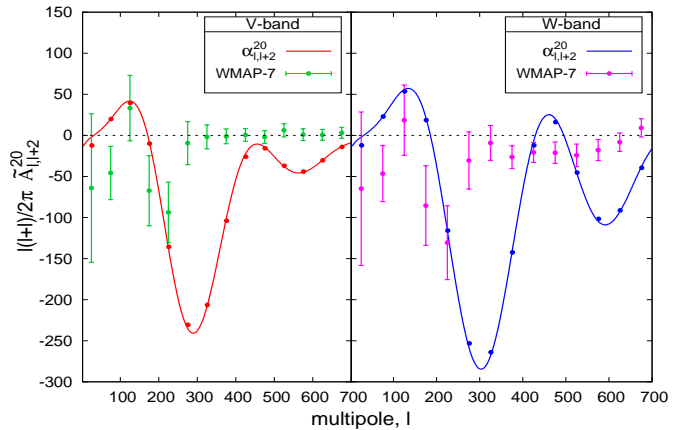}
\captionsetup{singlelinecheck=off,justification=raggedright}
\vskip 0.3cm
\captionof{figure}{The signal in the BipoSH coefficients
$A^{20}_{l,l+2}$ predicted due to lensing (See Eq. \ref{cute1} ) plotted
against the detected signal from WMAP 7 year maps. The value of the parameter
$\psi_{20}$ is set to the best fit value derived from the fit obtained to the
BipoSH coefficients $A^{20}_{ll}$ and can be found in Table \ref{table1}
.}
\label{fig2}
\end{figure}

In our analysis we use the best-fit $\Lambda$CDM CMB angular power spectrum
$C_l$ generated using the 
publicly available Boltzmann code \href{http://www.camb.info}{CAMB} \cite{CAMB}. We use the
lensing field harmonic coefficient $\psi_{20}$ which appears in Eq. \ref{cute}
as a free parameter, as currently no measurements of the projected lensing
potential exist at the largest angular scales. We perform a simple $\chi^2$
minimization technique to estimate the bestfit value of this free parameter.
Note that we carry out the fitting procedure only with the BipoSH coefficients
$A^{20}_{ll}$, since the detections in these coefficients are highly
significant. Our analysis yields a value of $\psi_{20}=2.25\times10^{-2}$ with
which we are
able to fit the detected signal in the BipoSH coefficients $A^{20}_{ll}$ in both
V-band and W-band maps equally well, after accounting for the effects of the
respective instrumental beams. Since the harmonic coefficients are coordinate
dependent quantities, we specifically mention that this value of $\psi_{20}$
inferred is in ecliptic coordinates.
\begin{table}[!h]
\setlength{\tabcolsep}{3pt}
\renewcommand{\arraystretch}{1.3}
\begin{center}
\begin{tabular}{|c|c|c|c|}
\hline
& FWHM (in degrees) & $\chi^2$ per d.o.f& $\psi_{20}$ \\
\hline
V-band & $0.326$ &  $0.818$ & \multirow{2}{*}{$2.25\times10^{-2}$}
\tabularnewline
\cline{1-3}
W-band & $0.202$ & $0.854$ & \tabularnewline
\hline
\end{tabular}
\end{center}
\vskip -0.3 cm
\captionsetup{singlelinecheck=off,justification=raggedright}
\captionof{table}{WMAP beam specifications from
\href{www.lambda.gsfc.nasa.gov}{LAMBDA} site. The $\chi^2$ per degree of freedom
(d.o.f) for the best-fit $\psi_{20}$ parameter that is obtained
by fitting Eq. \ref{cute} to the BipoSH coefficient $A^{20}_{ll}$
detections in WMAP 7 year maps.}
\label{table1}
\end{table}
The results of the whole analysis are summarized in Table \ref{table1} and
Figure \ref{fig1}.
The exceedingly good fit to the data as suggested by the reduced $\chi^2$ values
in the above table are due to over estimates of the standard deviation on the
data points in absence of inverse noise weighting. The best fit value of the parameter $\psi_{20}$ is then used to
predict the signal due to lensing in the BipoSH coefficients $A^{20}_{l,l+2}$
(See Eq. \ref{cute1} ). The results are plotted against the detections found
from WMAP 7 year maps and are depicted in Figure \ref{fig2}. In this case the
predicted signal does not seem to match the detections, however this could be
because of other systematic effects which remain unaccounted, like the
non-circular beam. Nevertheless it is interesting that the predicted signal for
these BipoSH coefficients follows the trend seen in the detections particularly
at low multipoles. 

We have clearly established that gravitational lensing can introduce significant
corrections to the BipoSH coefficients rendering them non-zero even in an
isotropic universe. We have also demonstrated that the difference in amplitude
of the signal detected from V-band and W-band maps can be explained by
accounting for the respective beam angular sensitivities. 

The best fit $\Lambda$CDM cosmology predicts the quadrupolar power in the
projected lensing potential power spectrum $C_2^{\psi \psi}$ to be of the order
of $\sim 10^{-8}$ requiring the corresponding harmonic coefficients $\psi_{2m}$
to be of order $\sim 10^{-4}$. This suggest that the value of the quadrupole
estimated in our analysis would be highly improbable.
Hence in the realm of standard $\Lambda$CDM cosmology the observed signal in the
BipoSH coefficients may not be completely explained by accounting for
modifications due to lensing. Hanson et. al. \cite{hanson-2010-81} have argued
that the detections found in the BipoSH coefficients may be explained by
accounting for the non-circular nature of the WMAP instrumental beam. The
detections of the BipoSH coefficients in the WMAP 7 year data may be completely
explained by accommodating the total contribution from both these effects.

After correcting for the systematic effects of the beam, it will be important to
account for the lensing modifications to the CMB two point correlation function
as has been discussed throughout this article. It is also pertinent to note that odd parity BipoSH of HS that may be associated with the effect non-circular beam with a specific scan strategy could possibly distinguish between the two effects. 
 Otherwise, this would require a completely
independent measurement of the lensing potential field and can be only provided
by LSS measurements. In the following section we discuss the LSS measurements
that will be essential to estimate the projected lensing potential $\Psi$.

The projected lensing potential is constructed by forming a linear weighted sum
of the gravitational potentials along the line of sight. Since the dominant
contribution to the gravitational potentials is due to dark matter, an all sky
map of the large scale distribution of dark matter will be needed. This will be
made possible through gravitational lensing surveys. A significant contribution
to the projected lensing potential power spectra comes from LSS below redshift
$z\lesssim5$ \cite{AL-AC}. However to estimate the low multipole moments of the
projected lensing potential, it will suffice to map the dark matter distribution
upto redshift $z\lesssim 2$. This will allow a reasonable estimate of the
lensing contribution to the detections in the BipoSH coefficients.

The Cosmic Evolution Survey \href{http://cosmos.astro.caltech.edu}{COSMOS} has
mapped the dark matter distribution out upto redshift $z=5$, however only on a
small patch of the sky covering 2 square degrees. Similar deep surveys with much
larger sky coverage will be required to achieve this goal. This will be made
partially possible with upcoming surveys like Large Synoptic Survey Telescope
\href{http://www.lsst.org}{LSST}, Dark Energy Survey
\href{http://www.darkenergysurvey.org}{DES}  and
\href{http://www.sci.esa.int/science-e/www/area/index.cfm?fareaid=102}{EUCLID}.

Next, we discuss the possibility of completely explaining the detection of the
BipoSH coefficients, as arising from the lensing modifications alone. The
statistically significant detections of the quadrupolar bipolar power spectrum
in the recent WMAP 7 year results \cite{CB-RH-GH} can be explained by accounting
for the lensing corrections to the BipoSH coefficients. 
This explanation would imply that the large scale distribution of dark matter surrounding us
happens to have an extremely high quadrupole moment.

There have been other anomalies, like the cold spot observed in
the CMB maps, which may not be explained in the realm of the standard
$\Lambda$CDM cosmology. It has been suggested that this anomaly in the CMB maps
may be explained by correcting the observed CMB maps for integrated Sachs-Wolfe
(ISW) effect due to an immensely large void in the LSS surrounding us \cite{szapudi,subir}. Further
it has been argued that the presence of such large voids may only be explained
by invoking non-Gaussianity in the primordial density fluctuations \cite{AY-BW}.
The best fit value of $\psi_{20}$ obtained in our analysis may not seem as
improbable if specific non-Gaussian initial conditions are invoked. More follow up work is motivated by our results. 

We have demonstrated that the difference in the amplitude of the signal detected
in WMAP V-band and W-band maps can be explained by accounting for the beam
angular sensitivities.  Further, Groeneboom et. al. \cite{groeneboom-2009} assess
that the non-circular beam cannot explain the non-zero detections of the BipoSH
coefficients, contesting the claims made by Hanson et. al. \cite{hanson-2010-81}.
These arguments seem to allow for the possibility of the
detection in the BipoSH coefficients having a cosmological origin. A weak violation of isotropy may result in a relatively large value for the quadrupole of the projected lensing potential which in effect could magnify the SI violating signal through lensing. In such a case the detections may actually suggest a violation of statistical isotropy \cite{ackerman,pullen,MA-TS,DP-DB-TS}.

Finally, the value of the parameter $\psi_{20}$ inferred could be interpreted as
first explicit measurement of projected lensing potential at the largest angular
scales. Gravitational lensing has been proposed as one of the tests for modified
gravity models \cite{schmidt}, suggesting yet another possibility of constraining these models through
measurements of the lensed BipoSH coefficients.

We acknowledge the use of HEALPix package \cite{healpix}. We acknowledge useful discussions
with Tuhin Ghosh and Nidhi Joshi. AR acknowledges the Council of Scientific and
Industrial Research (CSIR), India for financial support (Grant award no.
20-6/2008(II)E.U.-IV). MA and TS acknowledges support from the Swarnajayanti
fellowship, DST, India.
\bibliography{ref}
\bibliographystyle{apsrev}
\end{document}